\documentclass{article}

\usepackage[final, nonatbib]{bdl_2018}

\usepackage[utf8]{inputenc} 
\usepackage[T1]{fontenc}    
\usepackage{hyperref}       
\usepackage{url}            
\usepackage{booktabs}       
\usepackage{amsfonts}       
\usepackage{nicefrac}       
\usepackage{microtype}      
\usepackage{xcolor}
\usepackage{bm}
\usepackage{amsmath}
\usepackage{graphicx}
\usepackage{amsmath}
\def\rvp{{\mathbf{p}}}

\title{BDLOB: Bayesian Deep Convolutional Neural Networks for Limit Order Books}

\author{
  Zihao~Zhang, Stefan~Zohren, Stephen~Roberts\\
  Department of Engineering Science, Oxford-Man Institute of Quantitative Finance,
  University of Oxford\\
  \texttt{ \{zihao,~zohren,~sjrob\}@robots.ox.ac.uk}
}

\begin{document}

\maketitle

\begin{abstract}
We showcase how dropout variational inference can be applied to a large-scale deep learning model that predicts price movements from limit order books (LOBs), the canonical data source representing trading and pricing movements. We demonstrate that uncertainty information derived from posterior predictive distributions can be utilised for position sizing, avoiding unnecessary trades and improving profits. Further, we test our models by using millions of observations across several instruments and markets from the London Stock Exchange. Our results suggest that those Bayesian techniques not only deliver uncertainty information that can be used for trading but also improve predictive performance as stochastic regularisers. To the best of our knowledge, we are the first to apply Bayesian networks to LOBs.
\end{abstract}

\section{Introduction}
\label{introduction}
Electronic Limit Order Books (LOBs) \cite{parlour2008limit} are commonly used in modern financial markets. We can consider LOBs as the most granular financial data that records all limit orders\footnote{Limit orders are resting order to either buy or sell an instrument, i.e. those that do not match immediately. All resting limit orders form the LOB which represents the supply and demand of a given financial instrument at any moment in time.} of an exchange. Limit order data, like any other financial time-series data, is generally non-stationary and has a notoriously low signal-to-noise ratio.

Many models used historically for LOB data, such as the vector autoregressive model (VAR) \cite{zivot2006vector}, operate on handcrafted features extracted of the data to improve predictability. It is far from trivial to select informative features by hand. Recent research \cite{tsantekidis2017forecasting, sirignano2018universal, zhang2018deeplob} has shown that deep learning models can deliver good predictive performance using only raw LOBs data with the process of feature extraction being automated by convolutional layers. However, such deep learning techniques are (typically) non-probabilistic and thus provide only heuristic uncertainty measures associated with output variables. For any decision making process, especially for high-frequency trading, trust and risk are primary concerns.

Uncertainty may be quantified through Bayesian inference. Given the complexity of network models, such \emph{Bayesian Neural Networks} \cite{bishop1995neural} are often achieved by approximation such as variational inference \cite{bayes2008variational}. The work in \cite{gal2016dropout} proposed \emph{dropout variational inference}, also known as dropout sampling, as an approximation to BNNs. While variational dropout has been well established and deployed to many application domains, we have not found any literature applying it to LOBs.

\emph{Our contributions:} We apply variational dropout to a large-scale deep learning model designed in \cite{zhang2018deeplob} to predict short term movements in cash equity prices from LOBs. We showcase that dropout techniques, as stochastic regularizers, achieve better predictive performance and we design trading strategies that can deliver improved returns using derived uncertainty information. Further, we show how uncertainty relates to position sizing, avoiding bad trades and amplifying profits.

\section{Methods}
\label{methods}
We here briefly introduce the dropout variational inference \cite{gal2016uncertainty} and the network architecture, known as DeepLOB, in \cite{zhang2018deeplob}. In addition, we discuss different trading strategies and demonstrate how uncertainty information derived from Bayesian networks can be used to decide entry and exit points and further to upsize or downsize our positions. 

\subsection{Dropout Variational Inference and Network Architecture}

For Bayesian networks that have more than one hidden layer, the true posterior distributions of model parameters can not usually be evaluated analytically. Instead, we can define a variational distribution, which has a simple structure to evaluate, to approximate the true posterior distribution. Following this variational interpretation, the work of \cite{gal2016uncertainty} showed that dropout can be considered as an approximating distribution to the posterior in a Bayesian network. In other words, the weights obtained from the optimisation of a network using dropout are the same as the variational parameters of a network that has the same structure. We can then model uncertainty in Bayesian networks by a technique called \textit{Monte-Carlo (MC) dropout}, which is essentially running a number of stochastic forward passes through our network and averaging the results, 

The network architecture in \cite{zhang2018deeplob} is composed of three main building blocks: convolutional layers, an Inception Module and LSTMs. It is often difficult to extract representative features in financial applications as there are enormous observations and financial data is notoriously noisy. Hand-engineered features, such as technical indicators or PCA, often rely on tacit assumptions that may not well describe financial data and the process is non-trivial. The usage of CNNs and Inception Modules is then to automate the process of feature extraction, removing the above constraints. Feature maps obtained from convolutional layers are then passed into a LSTM layer instead of fully connected layers since these features still contain time dependencies. We can use LSTMs to capture temporal relationship among the resulting feature maps. 

Dropout was not used in the above work but we apply it to obtain uncertainty estimates. However, in order to get well-calibrated uncertainty, we need to adapt the dropout probability and this process is often time consuming for the large model and dataset. As a result, we apply a dropout layer with rate 0.2 after the Inception Module, decided using grid-search methods. Note that the feature maps are still time-series, so we keep the dropout mask the same for all timesteps. Doing this would not break time dependencies but regularize resulting features. Further, we applied Concrete Dropout \cite{gal2017concrete} to convolutional layers in the Inception Module and found that all dropout rates are pushed towards zero due to the large number of observations we have available. Further, we note that the convolutional layers have considerably less parameters compared to the dense layers because of the parameter sharing structure, so they are less likely to be overfitted \cite{wu2015towards}. 


\subsection{Different Trading Strategies}

The work of \cite{zhang2018deeplob} classified future price movements into three classes (up=+1, neutral=0, down=-1) and used a softmax layer to give a probability vector at each time stamp $\rvp_t = [p_{+1,t}, p_{0,t}, p_{-1,t}]^T$. The basic trading strategy designed in \cite{zhang2018deeplob}  to test the quality of the prediction was based on discrete outputs and did not use the softmax probability vector. We briefly describe this strategy below and refer to it as \textbf{Normal trading strategy}. We then show how returns can be amplified by incorporating information from the softmax layer (\textbf{Softmax trading strategy}) and finally show how positions can be sized by using uncertainty information from the Bayesian networks (\textbf{Bayesian trading strategy}).

\paragraph{Normal trading strategy}

The model delivers a softmax probability vector, $\rvp_t$, at each time. We enter the long position if $max(\rvp_t) = p_{+1,t}$ or the short position if $max(\rvp_t) = p_{-1,t}$. Only one position is allowed and we hold until the opposite movement is predicted (i.e. $max(\rvp_{t+k}) = p_{-1,t+k}$ in case of a long position where $k$ is the holding period). We go long or short (buy or sell) the same number of shares ($\mu$) at each time and set $\mu$ to be 30\% of first ask or bid level depending on the position we enter.

\paragraph{Softmax trading strategy}

Same as above, but we use softmax probability vector to constrain our timing of entering positions. Instead of entering the trades when $max(\rvp_t)$ is equal to $p_{+1,t}$ or $p_{-1,t}$, we place a simple rule to only enter the trade if $max(\rvp_t)$ is greater than a threshold $\alpha$. We place no constraint to exit the trade. 

\paragraph{Bayesian trading strategy}

While we use softmax outputs in the previous strategy, the network outputs could be biased and it is often argued \cite{gal2016uncertainty} that softmax probability vectors can not be interpreted as true model confidences. Instead, three methods, variation ratios, predictive entropy and mutual information are suggested in \cite{gal2016uncertainty} to summarise uncertainty in classification when using variational dropout. We tested on all three quantities and discover that predictive entropy, $\mathbb{H}$, gives most consistent uncertainty information and we use this information to size our positions. For a given input $\bm{x}$, we run $M(=100)$ stochasitc forward passes through our network and average the probabilities of each class to obtain the averaged probability vector $\bm{\bar{\rvp}}_t$. As before, we still enter trades if $max(\bar{\rvp}_t) > \alpha$, but we upsize our positions to $1.5 \times \mu$  if $\mathbb{H}<\beta_1$ or keep original size $\mu$ if $\beta_1<\mathbb{H}<\beta_2$ but downsize our positions to $0.5 \times \mu$ if $\mathbb{H}>\beta_2$. We exit the position if $\mathbb{H} < \beta_2$. In next section, we show  how these trading parameters affect our profits and risk level. 


\section{Results}
\label{results}
The data we use are limit order book data for five stocks: Lloyds Bank, Barclays, Tesco, BT and Vodafone listed on London Stock Exchange for the entire 2017 year, totalling more than 134 million market quotes. We take the first 6 months as training data, the next 3 months as validation data and the last 3 months as test data. Our test data contains millions of observations that can thoroughly verify our model performance. We follow \cite{zhang2018deeplob} to prepare our input data.

\begin{figure}[h]
  \centering
  \includegraphics[width=5.5in, height=2.5in]{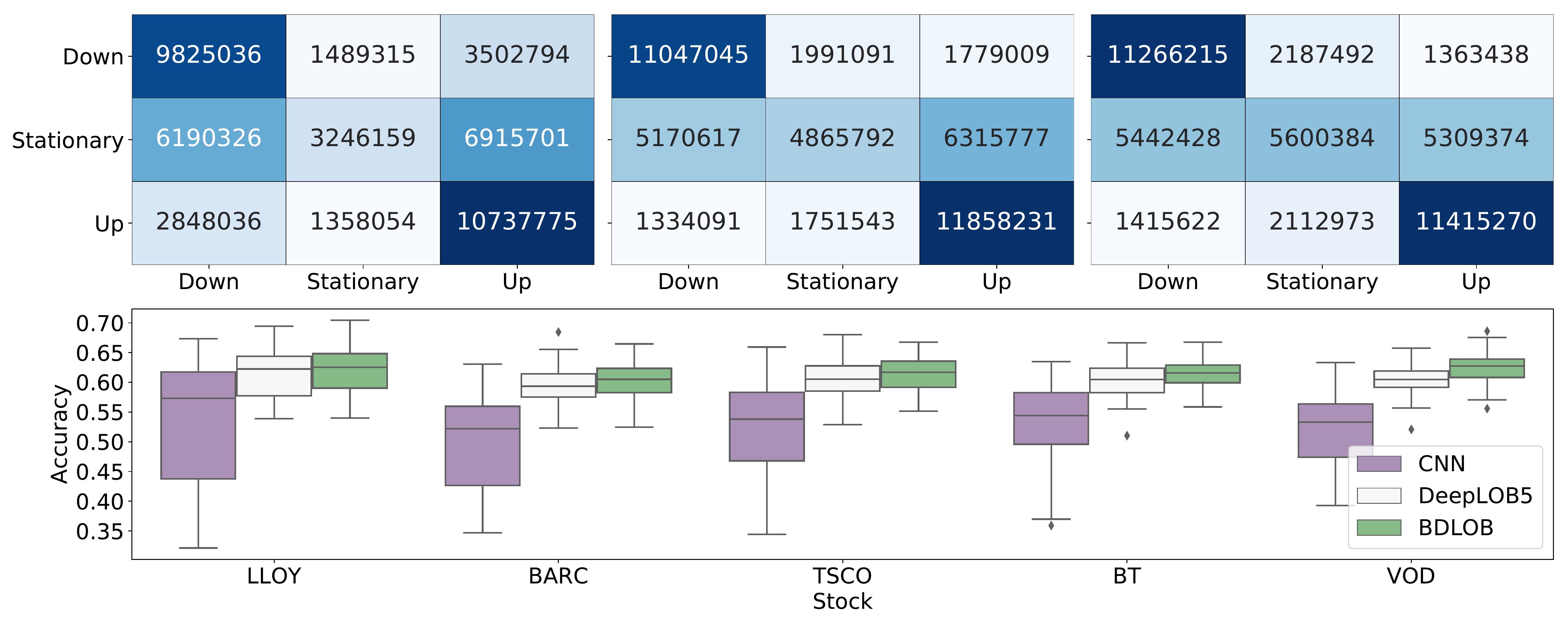}
  \caption{\textbf{Top:} confusion matrices for CNN, DeepLOB5 and BDLOB; \textbf{Bottom:} boxplots of daily accuracy for the different models across 5 stocks.}
  \label{fig:cm_daily_accuracy}
\end{figure}

We denote our model as BDLOB and compare it to two baseline models: the CNN model presented in \cite{tsantekidis2017forecasting} and DeepLOB5 in \cite{zhang2018deeplob}. We show each model's precision, recall, F1 score and average AUC score in Table~\ref{table:metrics} and we present confusion matrices and boxplots for daily accuracy in Figure~\ref{fig:cm_daily_accuracy}. BDLOB achieves the best results since dropout, as stochastic regularizers, improves generalization ability. Also, from Figure~\ref{fig:cm_daily_accuracy}, we can see that BDLOB has better ability to classify stationary class and lead to consistent better performance with narrow interquartile range (IQR) and less outliers.

\begin{table}[h]
  \caption{Evaluation metrics for different baseline models}
  \label{table:metrics}
  \centering
  \begin{tabular}{l|llll}
    \toprule
    Model     						& Precision     & Recall  &  F1 Score & AUC \\
    \midrule
    CNN   								& 0.52   & 0.52  	 & 0.48    & 0.672\\
    DeepLOB5   					& 0.60  	& 0.60    & 0.58  & 0.803\\
    BDLOB   & 0.60  	& 0.61    & 0.60  & 0.811\\ 
    \bottomrule
  \end{tabular}
\end{table}

We follow the same assumptions, mid-price simulation and no transaction costs, as in \cite{zhang2018deeplob} and present how different trading strategies affect profitability. Our focus is not to design a stand-alone trading strategy but to showcase how uncertainty information can be beneficial to decide enter and exit points and also position sizing. However, due to the short holding periods, most high-frequency trading firms are engaged in market making. Being able to trade passively on one side of the trade effectively corresponds to a mid-mid trade, being able to trade both sides passively reduces the fee further. The above, simplified trading simulation still approximates the relative value of the here presented model when added to an existing market making strategy. 

\begin{figure}[h]
  \centering
  \includegraphics[width=5.3in, height=2.9in]{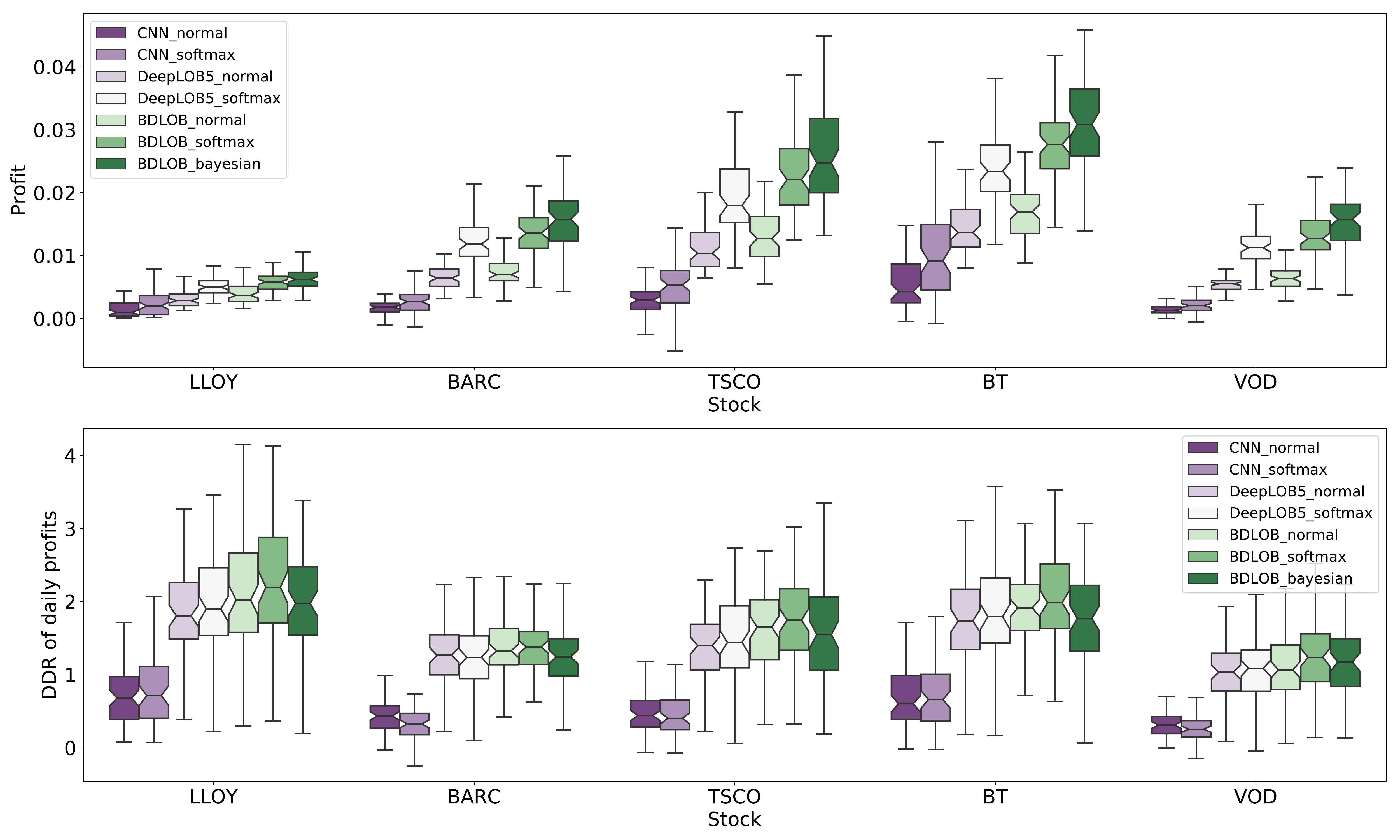}
  \caption{Boxplots and DDR for normalized daily profits. Profits are in GBX(=GBP/100).}
  \label{fig:boxplot_return}
\end{figure}

\begin{figure}[h]
  \centering
  \includegraphics[width=5.5in, height=2.6in]{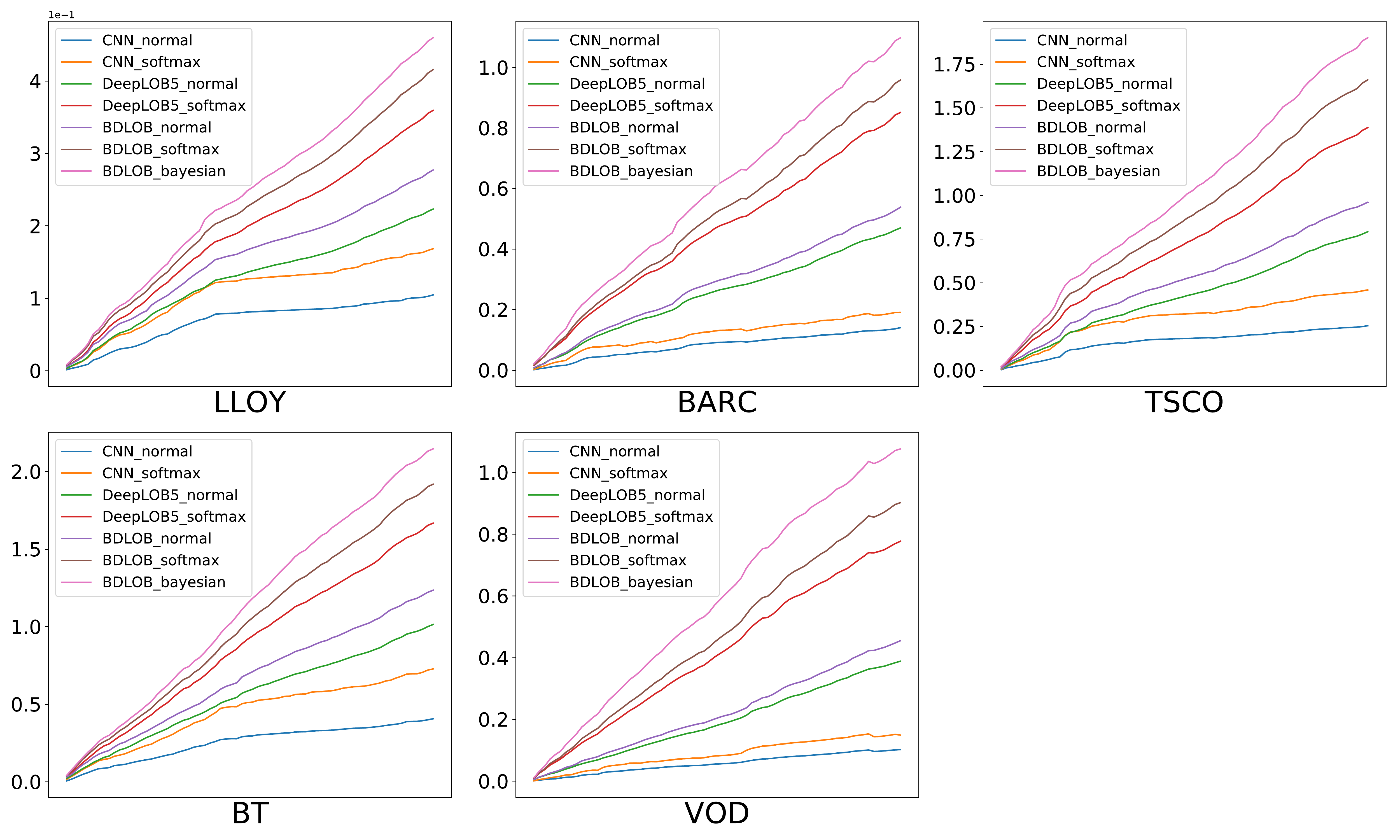}
  \caption{Cumulative profits for test periods for different models. Profits are in GBX(=GBP/100).}
  \label{fig:cum_return}
\end{figure}

We hold no stocks overnight and close all positions by the end of the day. As each strategy leads to different trading volumes, we normalise each day's return by total volumes executed on that day. In the literature, the Sharpe Ratio \cite{sharpe1994sharpe} is often used to measure a strategy's risk. However, the Sharpe Ratio is based on variance of returns that treats both positive and negative returns to be ``risky''. In practice, few traders would consider large positive returns as risky so we use the Downward Deviation Ratio (DDR) \cite{moody2001learning} as a measure of risk. The DDR penalizes  negative returns and rewards large positive returns. The DDR is defined as,
\begin{equation}
\mathrm{DDR} = \frac{E(R_t)}{\mathrm{DD}_{T}} \quad \text{where} \quad \mathrm{DD}_{T} = \Big(\frac{1}{T}  \sum_{t=1}^T \min(R_t, 0)^2 \Big)^{\frac{1}{2}}
\end{equation}
Figure~\ref{fig:boxplot_return} presents the boxplots for daily profits normalized by total volumes on that day. Any strategy using uncertainty information leads to substantial improvements in returns and BDLOB delivers the most profits and maintain a similar risk level.  Figure~\ref{fig:cum_return} shows the cumulative profits across testing periods and we observe consistent and better performance for BDLOB. We set $\alpha = 0.7$ for all strategies that use softmax probability vectors and set $\beta_1=0.1, \beta_2=0.9$ for Bayesian trading strategy. The value of these hyperparameters are selected to balance profits and risk. We discuss these choices below.

\begin{figure}[h]
  \centering
  \includegraphics[width=5.5in, height=3.9in]{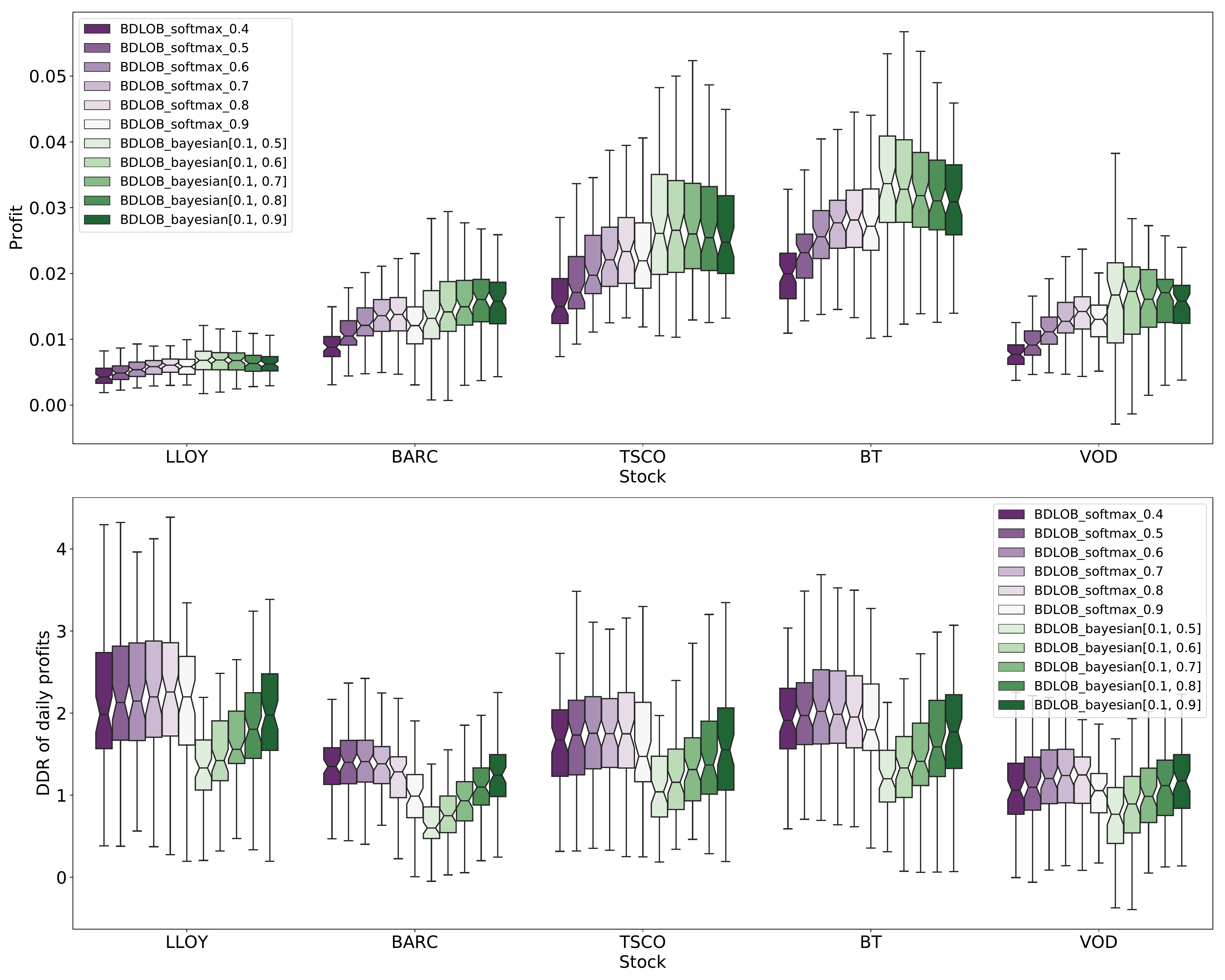}
  \caption{Boxplots for normalized profits and DDR for different trading parameters. We denote a softmax strategy with rate $\alpha$ as BDLOB\_softmax\_$\alpha$ and Bayesian strategy as BDLOB\_bayesian$[ \beta_1, \beta_2]$. We set $\alpha=0.7$ for all Bayesian trading strategies.  Profits are in GBX(=GBP/100).}
  \label{fig:boxplot_hyper}
\end{figure}

Figure~\ref{fig:boxplot_hyper} shows how profit and risk change according to these hyperparameters. We can see that a higher $\alpha$ generally improves profits but $\alpha=0.8$ or $\alpha=0.9$ can lead to larger risk or lower profits -- this is due to few trades taking place so any wrong prediction has a stronger impact on performance. As a result, we select $\alpha=0.7$ as the best value and apply it to all Bayesian trading strategies in Figure~\ref{fig:boxplot_hyper}. We set $\beta_1 = 0.1$ and find profit or risk is not hugely affected by this value because our model is very confident in its predictions when a high $\alpha=0.7$ is used. The effect of $\beta_2$ is obvious as it not only penalizes unnecessary trades but also controls the exit time point. We can see that high profits can be obtained with a smaller $\beta_2$ but with a much higher risk level. A strict exit rule allows us to take even fewer trades amplifying profits, but the impact of wrong predictions is also more obvious and we can hold a bad trade for too long without leaving it, leading to large variance.  

\section{Conclusion}
\label{conclusion}
We show that Dropout inference improves performance for predicting stock price movements and demonstrate that uncertainty information can be utilised to design better trading strategies. A promising direction for future work is to explore recent Bayesian techniques such as \cite{khan2018fast} and \cite{badam} and consider trading strategies that monitor uncertainty information during holding periods.


\end{document}